\def\year{2019}\relax
\documentclass[letterpaper]{article} 
\usepackage{aaai19}  
\usepackage{times}  
\usepackage{helvet} 
\usepackage{courier}  
\usepackage[hyphens]{url}  
\usepackage{graphicx} 
\usepackage{graphicx}  
\title{Priority Quality Attributes for Engineering AI-enabled Systems}
\author{Lena Pons, Ipek Ozkaya\textsuperscript{\rm 1} \\ 
\textsuperscript{\rm 1}Software Engineering Institute Carnegie Mellon University\\ 
4500 Fifth Ave.\\
Pittsburgh, PA 15213\\
{lepons, ozkaya}@sei.cmu.edu 
}
 
\begin{document}

\maketitle

\begin{abstract}
Deploying successful software-reliant systems that address their mission goals and user needs within cost, resource, and expected quality constraints require design trade-offs. 
These trade-offs dictate how systems are structured and how they behave and consequently can effectively be evolved and sustained. 
Software engineering practices address this challenge by centering system design and evolution around delivering key quality attributes, such as security, privacy, data centricity, sustainability, and explainability. 
These concerns are more urgent requirements for software-reliant systems that also include AI components due to the uncertainty introduced by data elements. 
Moreover, systems employed by the public sector exhibit unique design time and runtime challenges due to the regulatory nature of the domains. 
We assert that the quality attributes of security, privacy, data centricity, sustainability, and explainability pose new challenges to AI engineering and will drive the success of AI-enabled systems in the public sector.  
In this position paper, we enumerate with examples from healthcare domain concerns related to these requirements to mitigate barriers to architecting and fielding AI-enabled systems in the public sector. 
\end{abstract}

\noindent The exponential increases we have seen in computational power and machine learning in the past decade have resulted in huge investments in systems that aspire to exploit powers of artificial intelligence (AI) and machine learning (ML). 
We can talk about the application of ML algorithms to models of billions of parameters, and terabytes and petabytes of data. 
Decision-making has a whole new meaning when driven by these algorithms. 
However, irrational enthusiasm and confusion are nothing new to the software engineering industry. 

\indent In time the disparity between hype and reality diminishes. 
For example, we already better appreciate the centricity of data to AI-enabled systems, which include not only software and hardware components, but also AI algorithms, models, and data.  
However, we are still in a quest to better understand for which unique characteristics AI-enabled systems need to be designed. 
Existing experience coming from industry and research highlight that abstraction boundaries in system elements erode more quickly in AI-enabled systems, especially those containing an ML model \cite{sculley2015}. 
Unanticipated dependencies create attack surfaces for which the system may not be secured. When the rate of evolution of data, the trained model, and the deployed model are misaligned, sustaining AI systems creates costly outcomes, including but not limited to wrong inferences resulting in taking the systems out of production or creating unanticipated user consequences.  

\indent These emergent design characteristics in AI systems are exacerbated when coupled with the constraints of public sector systems. 
Barriers to deploying these systems in the public sector include: not understanding how a decision is made, no guarantee of security and privacy preservation, system misalignment with unanticipated data changes, and the introduction of new attack surfaces. 

\indent The robustness of AI-enabled systems will depend on how the system guards against failures related to these amplified properties. 
AI systems are software systems. Decision makers and developers of such systems in the public sector need to embrace AI engineering which accounts for building reliable systems by integrating data and software elements and uses design and analysis to manage the inherent uncertainty posed by algorithms  and other AI components. 
AI engineering addresses challenges such as inability to specify systems, managing hidden dependencies, and achieving verifiability with confidence. 
As more experience accumulates in deploying such systems a top priority outcome will be understanding what new architectural styles ensure the intended behavior with describable quantification of risk. 

\indent For software systems deployed in the public sector, quality attributes are not just principles that make for a better, more marketable product. 
They are the assurances required for a regulatory agency to make enforcement determinations about whether a software system complies with the legal and regulatory requirements for the context in which it is deployed. 
In this paper, we summarize unique characteristics of security, privacy, data centricity, sustainability, and explainability as they relate to architecting AI-enabled systems for the public sector. 
We drive our examples from healthcare and specifically discuss AI-enabled Software as a Medical Device (SaMD). 

\section{Quality Attributes and AI Engineering}

\noindent Quality attributes in software systems are defined as a property of a work product or goods by which its quality will be judged by some stakeholder or stakeholders \cite{ISO25010}. 
Quality attributes are and should be quantifiable in specifications by the definition of some appropriate and practical scale of measure \cite{clements2010}. 
Software systems meet their business and mission goals to the extend they meet their quality attribute requirements. 
AI systems, in particular those with ML elements, are systems that learn from data for decision making, hence are not designed to upfront requirement specifications. 
However, this assertion applies to the knowledge inference and modeling aspects. 
The overall quality concerns and expectations that should drive the expected behavior and quality of AI systems do still follow principles that can be known a priori, and evolved in conjunction with the AI-enabled systems. 

\indent The practice of software engineering for systems that incorporate an AI component is at present an open research topic. 
We identify four dimensions on which AI engineering most strongly needs to make immediate progress on: building robust systems, data, human-machine interaction and models.  
The quality attributes we highlight – security, privacy, data centricity, sustainability and explainability – illustrate how techniques for developing such systems as trustworthy robust software are required along these four dimensions. 
We discuss these attributes individually and identify cross-cutting concerns, particularly with respect to data.

\subsection{Security}
\noindent Securing AI-enabled systems amplifies the challenges associated with securing software systems to a level we are ill-equipped to address. 
AI systems have new attack surfaces. 
The learning element of some AI algorithms can provide a powerful vector for an attacker to change the output of a system, and in a software system where AI is just one component, the implications for the security of the system as a whole can be difficult or impossible to model or test. 
Additionally, the test space for an AI system explodes, when we are considering a system designed to handle billions of features \cite{breck2017}. 

\indent AI-enabled systems frequently rely on abstracting the input data into a data representation that is not human interpretable. 
One particularly difficult problem in securing AI systems is that the models and intermediate stages of an AI system are not comprehensible by humans without machine assistance. 
This makes fundamental aspects of verification and validation not tractable for AI systems. These challenges exist across many AI algorithms, but are particularly true of deep learning models, where the model is a black box. 

\indent For public sector software systems, this frequent inability to know that a system will perform as intended can create barriers to deployment from regulatory, risk management, and ethical perspectives. 
A software system that has substantial legal or financial ramifications for users or consumers of the products of that system ought to meet higher standards.  

\indent Securing AI systems requires new threat modeling of a system which can also impact how the systems is designed. 
The attack paths in these systems can be substantially more complicated, since changes to the data can translate into changes in the software system itself. 
Adversarial machine learning attacks such as adversarial inputs and data poisoning rely on the adversary taking advantage of the openness of the input space for an AI system. 
An approach like fuzzing for testing cannot be used to ensure the security of these systems.

\subsection{Privacy}

\noindent Privacy in software relates to the ability of individuals to have control and freedom of choice about the collection, use and disclosure of information about themselves. 
It should be clear to the users what information they are disclosing and providing owners of the data its control. There are existing guidelines and principles to help engineers understand privacy by design principles \cite{danezis2015}.

\indent Central AI concerns related to privacy include providing control to the owners of the data about what is shared, ensuring that derivative inferences are explained to the owners of the data and the AI systems are designed with such boundaries in mind. 
Design approaches should center around decoupling collection of data from its analysis in addition to mechanisms for securing.  

\indent Adversarial machine learning is implicated in privacy as well. 
A model stealing attack seeks to reverse engineer a model or its training data. 
This type of attack can expose individuals’ information if it is contained within the training data. 
And even if the training data alone is not sufficient to identify or deanonymize individuals, the potential for deanonymization from combining data from multiple sources is high \cite{benitez2010}.

\indent Advances in AI and ML software have been built on large volumes of data that have frequently been collected under the auspices that the information contained within that data can be anonymized. 
However, research has shown that unicity and re-identifiability in large data collections is frequently possible \cite{kondor2017}. 
In the case of many healthcare applications, the risk that unicity might exist in a data collection is enough to dissuade the development of applications, hindering research.

\subsection{Data centricity}

\noindent The role of data is the lynchpin aspect that makes engineering AI systems challenging, influencing every aspect of the system design. 
Software elements need to be architected for how data is structured, and how it behaves and need to be explicitly architected with the uncertainty, availability, and scalability of data in mind. 

\indent Most AI systems are centered around some ML model that has been trained on some data which is the weakest link. 
If the data presented to a trained model changes, then the performance of the system may degrade. 
A model that is designed to continuously learn from newly presented data is very fragile to adversarial attack, since the results of a model can be manipulated by presenting it with misleading or adversarial data. 
A trained model may produce results that cannot be critically evaluated by an expert because the model relies on patterns that are incomprehensible to the human expert. 
In fact, this property of an AI system is exactly the power of such a system. 
The interplay of a model, the data on which it was trained, and the data on which it makes predictions is the source of many of the difficulties with fielding trustworthy AI systems. 

\indent The scale of data poses its own set of engineering challenges. In the service of advancing AI techniques, entirely new data storage and database architectures have been developed. 
The permissiveness of data typing means that ensuring data quality can be a herculean task, which in turn adds to the attack surface of such systems.

\subsection{Sustainability}

\noindent A key challenge in AI systems is the different rates of change that may or may not be in sync with when the software that contains the AI/ML components are deployed, evolved, and replaced. 
A key principle of sustainability in AI systems is that changing anything changes everything.  
At the heart of these systems is a trained model. 
Even for algorithms that can deliver deterministic output, if the observed data that is presented to a model changes or the model is presented with something novel, then the performance of that trained model can decline substantially and unpredictably. 

\indent In the case of SaMD, sustainability includes striking a balance between improving the quality of deployed systems with maintaining the safety of such systems to be used in a clinical setting. 
For example, incorporating improvements to a model through continuous learning could improve an AI-enabled SaMD, but before the improvements could be pushed to the clinic, a series of tests would be required to re-affirm the validity. 
Monitoring performance can create substantial overhead and burden on human users, since providing validation data can range from a time-consuming annoyance in some domains to an impossibility in others.

\indent The sustainability challenge for SaMD is determining the pace at which to update the system. 
Retraining a model in an AI/ML system can range in terms of disruption to the system. 
And unlike a development cycle where updates to the system are planned, developed and tested on an external schedule, what triggers the need to refresh an AI system cannot always be determined by user needs. 

\indent The FDA recognizes this quandary. 
The agency has previously approved AI-enabled systems that preclude changing a deployed model while admitting that the power of these AI/ML-based SaMD lies within the ability to continuously learn. 
The agency feels comfortable fielding systems with a recognizable sustainability model. 
At the same time, the agency cannot resolve regulatory barriers to deploying more innovative methods to the clinic because the risks of continuously evolving and deploying are unpredictable and it would not be responsible to do so.

\subsection{Explainability}

\noindent The role of human machine teaming in AI systems is a paramount concern. 
The trustworthiness of an AI system relies on a user’s ability to understand how an algorithm has arrived at a result. 
Constructing explainable systems is a fundamental research question of AI. 
One tension in engineering AI systems is AI systems enable analysis that cannot be otherwise conducted, but these methods do not always permit the analyst to trace what feature(s) resulted in a prediction. 
When there are legal and ethical ramifications that arise from this tension, the frequent answer is to block such systems from being used in real scenarios.

\indent In SaMD systems, explainability is a fundamental concern of deploying such systems in the clinic. 
From the clinician’s perspective, the software system providing some insight into a diagnosis or treatment option must provide sufficient information for her to feel confident that the treatment option selected for the patient is the right one. 
From the patient’s perspective, he will not feel confident about the clinician’s recommendation if she cannot provide a reasonable explanation for why she chose a treatment option. 

\indent Developing techniques for explaining how outputs arise from AI systems to allow for critical review by experts will be a significant enabler to field these systems. 
Without such techniques, the public sector will continue to lag behind commercial capability from AI systems due to an inability to tolerate that level of risk. 
Designing a system such that information about intermediate stages of a prediction are stored and available to the user can be one way to aid in the explainability problem. 
Some algorithms can provide a measure of feature importance, which tells the user what features within a dataset contributed to an output. 
Storing and presenting intermediate output may help a user understand what information was used to arrive at a prediction. 
However, in the case of deep learning algorithms, the problem of explainability remains an evolving research area \cite{holzinger2018}. 

\section{Role of Architecting in AI Engineering}

\noindent An AI system must frequently be developed through the interplay of experts from computer science, software engineering, data engineering, statistics, machine learning and some other domain expertise. 
While teams of researchers have been working to develop AI software systems across multiple disciplines, basic ambiguities arise from differences in how experts communicate and understand the same problem. 
When you open the aperture to public sector software projects, the communication and conceptual challenges for understanding how AI systems interact with regulatory frameworks that were not designed with software in mind, much less AI-enabled software, the problem deepens. 
Government agencies are struggling to define how AI systems integrate into current regulatory regimes. 
At the same time, agencies are trying to determine the need for new regulatory authority and guidance to manage harms from these systems. 

\indent The drive toward AI engineering practices informed by quality attributes is fundamentally motivated by providing the kind of trustworthy and predictable outcomes desired by regulatory agencies. 
Additionally, regulatory agencies have bodies of expertise in administrative law, regulatory process, and subject matter expertise relevant to the regulated industry, but not frequently in statistics, machine learning, computer science or software engineering. 
These disciplines do not communicate naturally with one another and the depth of understanding in more than one domain required to effectively regulate it is rare and difficult to obtain.
\indent The motive for the discussion above driving quality attributes is to resolve this very tension inherent in fielding an AI system from these multiple stakeholder perspectives in a safety critical environment whose performance cannot be proven. 
Software architecture assists in this communication challenge by modeling systems such that they can be analyzed from multiple perspectives and for multiple purposes. 
Quality attributes drive this process \cite{clements2010}. 

\indent Security and privacy in AI systems are deeply enmeshed in the data that is used in the system. 
Researchers from every domain are seeking ways to leverage the power of AI algorithms without exposing sensitive data. 
Ability to achieve this also impacts the design of such systems, implying better ways of decoupling algorithms and data, hence improving ability to design for security, privacy, and sustainability. 
For public sector software systems, the regulatory requirements for handling, storing and sharing information are strict. 
As a consequence, researchers` ability to even discover what an AI application could do for many healthcare applications has been hindered by the regulatory requirements. 

\indent Software architecture provides a mechanism for analyzing dependencies to elucidate vulnerabilities and understand the attack surface of a software system. 
For systems that contain AI components, existing attack models may be insufficient to expose the risks that arise from data centricity and adversarial machine learning. 
For these complex systems that integrate AI components, new software architecture principles and patterns will need to be developed to capture the new risks to the system from the AI components.

\indent Abstraction and separation of concerns in software elements have been driving design and architecting principles with an intend to aid in deploying secure and quality software. 
The central role of data in AI systems challenges these principles, substantially compromising security and privacy. 
Our known architectural design tactics also do not suffice to  mitigate the risks either at an architectural or algorithmic level. 
Many of the privacy-preserving techniques that have been proposed, such as homomorphic encryption have proven too computationally expensive to implement in the context of AI systems and remain the purview of mathematicians and cryptographers. 

\section{Conclusion}

\noindent FDA published a discussion paper on a regulatory framework for AI/ML based software as a medical device that raises some of the inherent barriers that exist for fielding AI systems in safety critical environments. \cite{fda2019}. 
Specifically, FDA notes that many of the AI technologies that have been approved or cleared by FDA use of algorithms that are “locked,” which the authors define as “an algorithm that provides the same result each time the same input is applied and does not change with use.” 
Reading this definition in the context of AI engineering implies that FDA feels comfortable deploying an AI-enabled system provided that it is deterministic enough to fit within the software assurance frameworks of conventional software engineering. 

\indent This compromised approach that FDA suggests coupled with a disciplined focus on design and runtime attributes of AI-enabled systems as we discussed in this position paper offers a way ahead for the public sector while research and practices catches up. 
Repeatable practice will emerge as we understand more and create consistent vocabulary around expressing and analyzing for related design concerns. 

\section{Acknowledgments}
This material is based upon work funded and supported by the Department of Defense under Contract No. FA8702-15-D-0002 with Carnegie Mellon University for the operation of the Software Engineering Institute, a federally funded research and development center. The view, opinions, and/or findings contained in this material are those of the author(s) and should not be construed as an official Government position, policy, or decision, unless designated by other documentation. DM19-0845.

\bibliography{aaaiqual}
\bibliographystyle{aaai}

\end{document}